\documentclass[aps,twocolumn,showpacs,nobibnotes,epsf]{revtex4}

\usepackage{graphicx}
\usepackage{dcolumn}
\usepackage{bm}

\begin{document}
\title{Oxygen isotope effect on the superconductivity and stripe phase in La$_{1.6-x}$Nd$_{0.4}$Sr$_{x}$CuO$_4$}

\author{ G. Y. Wang, J. D. Zhang, R. L. Yang,        }
\author{X. H. Chen}
\altaffiliation{Corresponding author} \email{chenxh@ustc.edu.cn}

\affiliation{Hefei National Laboratory for Physical Science at
Microscale and Department of Physics, University of Science and
Technology of China, Hefei, Anhui 230026, People's Republic of
China\\ }

\date{\today}

\begin{abstract}
The oxygen isotope effect on the superconductivity, stripe phase
and structure transition is systematically investigated in
La$_{1.6-x}$Nd$_{0.4}$Sr$_{x}$CuO$_4$ with static stripe phase.
Substitution of $^{16}$O by $^{18}$O leads to a decrease in
superconducting transition temperature T$_C$, while enhances the
temperature of the structural transition from
low-temperature-orthorhombic (LTO) phase to
low-temperature-tetragonal (LTT) phase. Compared to the Nd free
sample, a larger isotope effect on $T_C$ is observed in
La$_{1.6-x}$Nd$_{0.4}$Sr$_{x}$CuO$_4$. These results indicate that
the distortion of CuO$_2$ plane suppresses the superconductivity,
giving a direct evidence for the competing of stripe phase and
superconductivity because the distortion of CuO$_2$ plane enhances
the stripe phase.

\end{abstract}


\maketitle
\newpage

High critical temperature (T$_C$) superconductor is widely studied
since the discovery of La$_{2-x}$Ba$_{x}$CuO$_{4}$ in 1986.
However, the properties of these materials are hardly to be
explained by BCS theory which is successful in explaining the
properties of conventional superconductor. It's still an open
question for the mechanism of high temperature superconductivity
in cuprates. Among numerous theoretical models, stripe phase has
attracted considerable attention that the spin and charge in high
T$_C$ superconductors distributes inhomogeneous and forms
"stripe"\cite{Zaanen,Emery1,Emery2}. It was experimentally
observed by neutron scattering or other method in
La$_{2}$CuO$_4$-based system
\cite{Hayden,Yamada1,Yamada2,Tranquada1,Tranquada2,Qu,Lee1,Moodenbaugh,Homes}
and YBa$_2$Cu$_3$O$_y$\cite{Mook1,Mook2,Sharma}. It generally
appears that the fluctuating stripe promote to superconductivity,
but static stripe may suppress superconductivity\cite{Orenstein}.
However, there is evidence that it is local magnetic order rather
than charge-stripe order which is responsible for the anomalous
suppression of superconductivity\cite{Ichikawa}.

In La$_{2}$CuO$_4$ system, several structural phase transitions
occur with doping of alkaline-earth and rare-earth
metals\cite{Axe, Crawford1, Buchner, Dabrowski, Crawford2,
Ichikawa}. With decreasing temperature a transition from
high-temperature-tetragonal (HTT) phase to
low-temperature-orthorhombic (LTO) phase, then
low-temperature-tetragonal (LTT) phase (or $Pccn$ phase, depends
on the hole concentration) was observed. The LTO and LTT phases
involve distortion of CuO$_2$ planes due to the tilting of CuO$_6$
octahedras producing stripe pinning potential. Here we call the
temperature T$_{LT}$ at which the structure transition from LTO to
LTT occurs. In La$_{2-x-y}$Nd$_{y}$Sr$_{x}$CuO$_4$, the
substitution of La by Nd enhances the pinning potential, which
pins the stripe from fluctuating (Nd free sample) to
static\cite{Tranquada1,Tranquada2}. When $x$ = 1/8 there is an
anomalous suppression of superconductivity due to the stripe
phase. In La$_{1.6-x}$Nd$_{0.4}$Sr$_{x}$CuO$_4$, neutron
diffraction experiment shows that the charge ordering and
structural transition are essentially coincident for x=0.10 and
0.12\cite{Tranquada2,Ichikawa}; however, the change ordering
occurs significantly below the structural transition temperature
T$_{LT}$\cite{Ichikawa}. Several groups have investigated the
relationship among structural distortion, stripe phase and
superconductivity, using high-pressure to control the structure
transition and superconductivity\cite{Arumugam, Crawford2}. It was
found that the hydrostatic pressure lower than 5 GPa compresses
the CuO$_2$ planes, which weakens the pinning potential,
suppresses the LTT distortion and enhances the superconductivity.

It is well known that the isotope effect study is very important
in the conventional superconductors in which $\alpha$$_C$=-
dln$T$$_C$/dln$M$$_O$=0.5, and is the illation of BCS theory.
Although many believe that antiferromagnetism is important for
superconductivity, there has been renewed interest in possible
role of electron-lattice coupling\cite{Lanzara,Devereaux}.
Therefore, the research of isotope substitution is necessary to
study mechanism of high-$T_c$ superconductivity. Several oxygen
isotope substitution experiments have been done in
La$_{2-x}$Sr$_x$CuO$_4$ and La$_{2-x}$Ba$_x$CuO$_4$ with
fluctuating stripe phase, and a large isotope exponent
($\alpha$$_C \sim 1$) on T$_C$ was found near 1/8
doping\cite{Crawford3, Crawford4, Zhao1, Zhao2}. To check possible
change in the isotope effect induced by Nd doping and investigate
the relationship among structural distortion, stripe phase and
superconductivity, we systematically study the oxygen isotope
effect in La$_{1.6-x}$Nd$_{0.4}$Sr$_{x}$CuO$_4$ with x=0.10,
0.125, 0.15 and 0.175. It is found that T$_C$ is suppressed, while
T$_{LT}$ is enhanced with the substitution of $^{16}$O by
$^{18}$O, indicating that the distortion of CuO$_2$ plane
suppresses superconductivity.  Because the charge ordering and
structural transition are essentially coincident for x=0.10 and
0.12, therefore, in this sense the results of isotope effect
definitely provide an evidence for the competing between stripe
phase and superconductivity. Compared to the Nd free sample with
fluctuating stripe phase, a larger isotope exponent $\alpha$$_C$
is observed in La$_{1.6-x}$Nd$_{0.4}$Sr$_{x}$CuO$_4$ with static
stripe phase, suggesting a strong electron-lattice coupling in
cuprates.

Polycrystalline samples La$_{1.6-x}$Nd$_{0.4}$Sr$_{x}$CuO$_4$ for
$x$ = 0.10, 0.125, 0.15 and 0.175 were prepared by conventional
solid-state reaction. All samples were characterized by X-ray
diffraction (XRD) and no observable impurity phase is found. One
pellet for each sample with different $x$ was cut into two pieces
for oxygen-isotope diffusion. The two pieces for each composition
were put into an alumina boat which were sealed in a quartz tube
filled with oxygen pressure of 1 bar (one for $^{16}$O$_{2}$ and
another for $^{18}$O$_{2}$) mounted in a furnace, respectively.
The quartz tubes formed parts of two identical closed loops. They
were first heated at 980 $^o$C for 90 h, then slowly cooled to 500
$^o$C, kept for 10 h and finally cooled to room temperature with
furnace. The obtained samples were re-examined by X-ray
diffraction to confirm them single phase. The oxygen-isotope
enrichment is determined by the weight changes of both $^{16}$O
and $^{18}$O samples. The $^{18}$O samples have about 80\%(
$\pm$5\%) $^{18}$O and 20\%( $\pm$5\%) $^{16}O$. To ensure the
isotope exchange effect, back-exchange of $^{18}$O sample by
$^{16}$O was carried out in the same way and the weight change
showed a complete back-exchange. Resistance measurements were
performed using the ac four-probe method with an ac resistance
bridge system (Linear Research Inc. LR-700P). To reduce the
experimental deviation, each couple of $^{16}$O and $^{18}$O
samples are measured synchronously in a cooling process.

\begin{figure}[t]
\centering
\includegraphics[width=8cm]{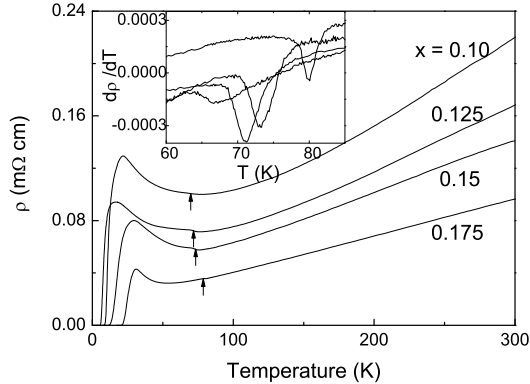}
\caption{Temperature dependence of resistivity for the samples
La$_{1.6-x}$Nd$_{0.4}$Sr$_{x}$CuO$_4$ with $x$ = 0.10, 0.125, 0.15
and 0.175 treated in $^{16}$O. The arrows indicate the LTO-LTT
transition temperature T$_{LT}$. Inset shows the derivative curves
of resistivity near T$_{LT}$.}\label{fig1}
\end{figure}

Temperature dependence of resistivity for the samples
La$_{1.6-x}$Nd$_{0.4}$Sr$_{x}$CuO$_4$ with $x$ = 0.10, 0.125, 0.15
and 0.175 treated in $^{16}$O are shown in Fig.1. T$_C$ (defined
as the midpoint of superconducting transition in resistivity) is
11K, 7.9K, 17.8K and 25K for $x$=0.10, 0.125, 0.15 and 0.175,
respectively, being consistent with that reported in other
literatures\cite{Buchner,Tranquada2,Ichikawa,Crawford2}. The
suppression of T$_C$ compared with
La$_{2-x}$Sr$_x$CuO$_4$\cite{Schneider} is attributed to the
static stripe phase induced by the substitution of Nd for La
atoms\cite{Tranquada1}. The abnormal suppression on T$_C$ near 1/8
doping can be clearly seen in our samples, that is, the $T_c$ is
the least for the sample with x=0.125. A small resistivity jump
appears at about 70K, which is indicated by an arrow and regarded
as the signal of structural transition from LTO to
LTT\cite{Crawford1, Ichikawa}. To show this jump clearly, the
temperature dependence of derivative of resistivity is shown in
the inset of Fig.1. A dip can be seen clearly in the derivative
curve. Here we define T$_{LT}$ as the dip temperature: 66.7K, 71K,
73.1K and 80K for $x$ = 0.10, 0.125, 0.15 and 0.175, respectively.
T$_{LT}$ increases with increasing Sr doping, consistent with that
reported in Ref.\cite{Ichikawa}.

\begin{figure}[t]
\centering
\includegraphics[width=8.5cm]{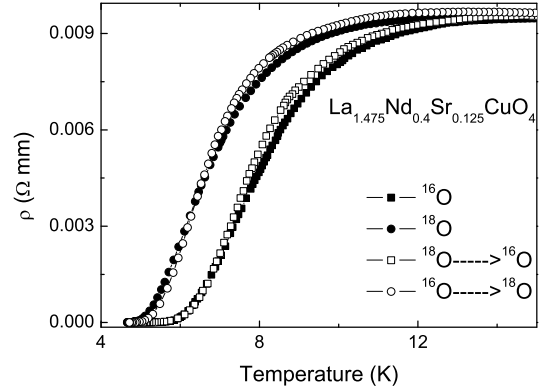}
\caption{Temperature dependence of resistivity near the
superconducting transition for the $^{16}$O and $^{18}$O samples
with $x$ = 0.125 .} \label{fig2}
\end{figure}

Figure 2 shows the temperature dependence of resistivity near the
superconducting transition for the sample with $x$ = 0.125. It
should be pointed out that the superconducting transition is a
little broadening, which may be caused by the fluctuation of Sr,
Nd and/or O contents. T$_C$ is 7.9K and 6.6K for $^{16}$O and
$^{18}$O samples, respectively. To ensure the change of T$_C$ from
isotope substitution, back-exchange of $^{18}$O sample by $^{16}$O
was performed. Fig.3 shows the Raman spectra at room temperature
for the sample with $x$ = 0.125. The apical O stretch mode is
softened from 433($\pm$1) cm$^{-1}$ to 413($\pm$1) cm$^{-1}$ by
the substitution of $^{18}$O for $^{16}$O. This frequency shift of
4.6\%$\pm$0.3\% suggests about 79\% $^{18}$O substitution because
Raman shift is in proportion to 1-$\sqrt{16/M'}$
\cite{Batlogg,Lee}.  The $^{18}$O substitution estimated by Raman
shift is consistent with the result obtained from weight change.
For comparison, $\rho$(T) of back-exchanged samples are also shown
in Fig.2. Two $^{16}$O/$^{18}$O samples show the same T$_C$, which
definitely indicates the change of $T_c$ arises from the oxygen
isotope exchange. The isotope exponent on T$_C$ in this sample
$\alpha$$_C$ = - dln$T$$_C$/dln$M$$_O$ is 1.89, much larger than
0.5 deduced from BCS theory.

\begin{figure}[t]
\centering
\includegraphics[width=8.5cm]{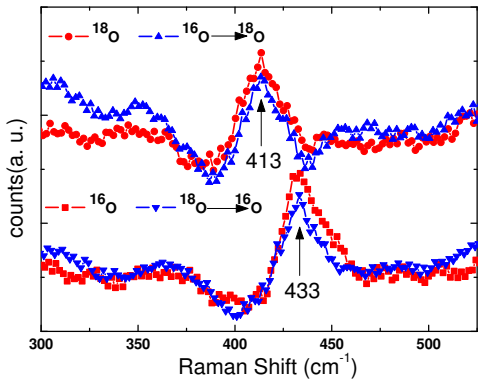}
\caption{Raman spectra at room temperature for the $^{16}$O and
$^{18}$O samples with $x$ = 0.125. A 514.5 nm Ar-laser line was
used as the excitation line.}\label{fig3}
\end{figure}

The phonon-mediated BCS theory shows that in condition of weak
electron-phonon coupling the increase in lattice mass enhances the
effective mass of charge carriers m$^\star$, and lowers the
superconducting gap $\Delta$, and finally suppresses T$_C$. This
used to be successful in explaining the isotope effect in most of
conventional superconductors, but failed in explaining the isotope
effect in high T$_{C}$ superconductor\cite{Schneider}. Especially
in La$_{2-x}$Sr$_{x}$CuO$_4$\cite{Zhao1,Zhao2,Crawford3,
Crawford4} the isotope exponent around 1/8 doping is about 1. Zhao
$et$ $al$.\cite{Zhao1,Zhao2} explain this with small polaron
theory that the effective mass of supercarriers depends strongly
on the oxygen-isotope mass in deeply underdoped regime, indicating
strong electron-phonon coupling in it. The isotope exponent
$\alpha$$_C \sim 1.89$ in La$_{1.6-x}$Nd$_{0.4}$Sr$_{x}$CuO$_4$
with x=0.125 is much larger than that ($\sim 1.0$) in  the Nd free
sample La$_{2-x}$Sr$_x$CuO$_4$. It indicates a stronger
electron-lattice coupling in
La$_{1.6-x}$Nd$_{0.4}$Sr$_{x}$CuO$_4$. Nd doping induces a
structural transition from LTO to LTT, which pins the stripe phase
from fluctuating (Nd free sample) to
static\cite{Tranquada1,Tranquada2} and enhances the distortion of
CuO$_2$ plane\cite{Buchner}. It has been reported in manganites
that lattice distortion tends to introduce stronger
electron-phonon coupling\cite{Zhao3}. Therefore, the electron
coupled to the more distorted CuO$_2$ plane induced by Nd doping
is responsible for the larger isotope exponent relative to the Nd
free sample.

\begin{figure}[t]
\centering
\includegraphics[width=8.5cm]{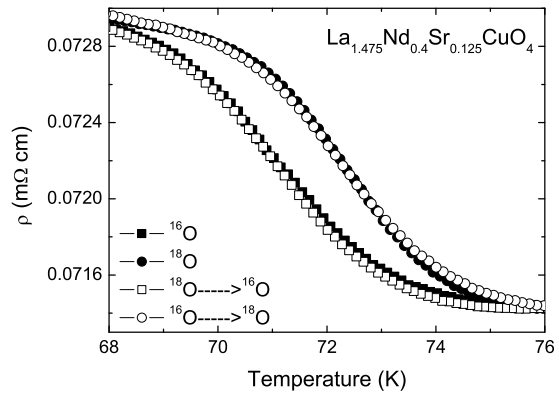}
\caption{Temperature dependence of resistivity near the structure
transition for the $^{16}$O and $^{18}$O samples of $x$ =
0.125.}\label{fig4}
\end{figure}

Figure 4 shows the temperature dependence of resistivity near
T$_{LT}$ for the samples of $x$ = 0.125. T$_{LT}$ is enhanced from
71K to 72.3K with the substitution of $^{16}$O by $^{18}$O,  the
isotope exponent $\alpha$$_{LT}$ is about -0.19. As shown in
Fig.4, the resistivities are almost the same for the
back-exchanged samples. It ensures that the change in T$_{LT}$
arises from the isotope effect. The increase of T$_{LT}$ indicates
the enhancement of stabilization for LTT phase, suggesting the
enhancement of distortion in CuO$_2$ plane by substitution of
$^{16}$O by $^{18}$O. As shown in Fig.2 and Fig.4, the
substitution of $^{16}$O by $^{18}$O leads to a decrease in $T_c$
and an increase in $T_{LT}$. The oxygen isotope effect provides an
evidence that the distortion of CuO$_2$ plane suppresses the
superconductivity, being consistent with the increase of $T_C$ by
lowering the impact of the disorder in Bi2212\cite{Eisaki}. It has
been reported that the charge ordering and structural transition
are essentially coincident for x=0.10 and
0.12\cite{Tranquada2,Ichikawa}. Therefore, the increase of
T$_{LT}$ indicates the enhancement of charge stripe, suggesting
the competing between the stripe phase and superconductivity.

For the samples La$_{1.6-x}$Nd$_{0.4}$Sr$_x$CuO$_4$ with x=0.10,
0.15 and 0.175, the oxygen isotope exponents for superconducting
transition $\alpha$$_C$ are 1.24, 0.98, 0.33, while for structural
transition $\alpha$$_{LT}$ are -0.32, -0.20, -0.17, respectively.
All samples show that the substitution of $^{16}$O by $^{18}$O
leads to a decrease in $T_c$ and an increase in $T_{LT}$.  Sr
content dependence of oxygen isotope exponent for superconducting
transition  $\alpha$$_C$ and  structural transition
$\alpha$$_{LT}$ is shown in Fig.5. The largest $\alpha$$_C$ is
observed in the sample with x=0.125, such 1/8 anomaly for
$\alpha$$_C$ has been reported in La$_2$CuO$_4$-based
superconductors\cite{Zhao1,Zhao2,Crawford3, Crawford4}. However,
the $\alpha$$_{LT}$ decreases with increasing Sr content. For
comparison, $\alpha$$_C$ reported in La$_{2-x}$Sr$_{x}$CuO$_{4}$
\cite{Franck,Zhao2,Batlogg,Crawford4,Hofer} is also shown in
Fig.5. It clearly shows a trend that $\alpha_c$ decreases with
increasing Sr content except for an anomaly around x=0.125 in
La$_{2-x}$Sr$_{x}$CuO$_{4}$. Our observation in
La$_{1.6-x}$Nd$_{0.4}$Sr$_x$CuO$_4$ shows similar trend except
that $\alpha$$_C$ for all Nd doped samples is larger than that in
La$_{2-x}$Sr$_{x}$CuO$_4$. It further indicates that the stronger
distortion of CuO$_2$ plane induced by Nd doping leads to a
stronger electron-lattice coupling. The sample with $x$ = 0.10
shows the largest $\alpha$$_{LT}$ value, which may be due to the
approach to the phase boundary from LTO to LTT/$Pccn$ for this Sr
content\cite{Ichikawa}. The large isotope effect caused by the
instability of lattice around phase boundary has also been found
in cobalt\cite{Wang}.

\begin{figure}[t]
\centering
\includegraphics[width=8.5cm]{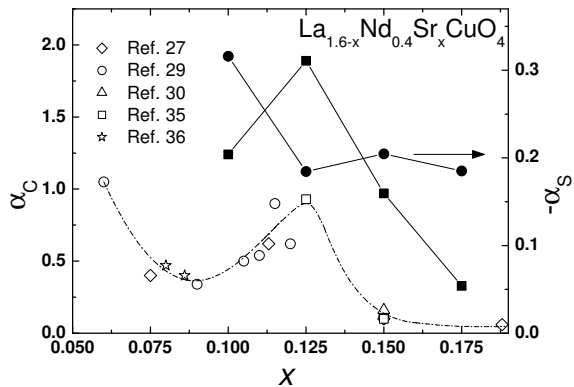}
\caption{$\alpha_c$ (solid squares) and $\alpha_s$ (solid circles)
as a function of x for La$_{1.6-x}$Nd$_{0.4}$Sr$_{x}$CuO$_4$. For
comparison, $\alpha_c$ from the previous works
 is also depicted for
polycrystalline and single crystals La$_{2-x}$Sr$_{x}$CuO$_{4}$.
The dash line guides eyes.}\label{fig5}
\end{figure}

In the phase diagram of La$_{2-x-y}$Nd$_{y}$Sr$_x$CuO$_4$ system,
T$_{LT}$ and the temperature of occurrence of stripe phase
increases simultaneously  with keeping Sr content constant and
increasing Nd content\cite{Ichikawa}. Therefore, it can be
believed that substitution of $^{16}$O by $^{18}$O leads to an
increase of $T_{LT}$, consequently to enhancement of stripe phase.
With keeping Nd content unchanged and increasing Sr doping level,
the temperature where stripe phase occurrs shows a hump as a
function of Sr doping level, while T$_{LT}$ increases with
increasing Sr doping level\cite{Ichikawa}. The suppression of
stripe phase for $x$$>$1/8 is just caused by the deviation of hole
concentration from 1/8. In our case, the oxygen isotope
substitution doesn't change the hole concentration. Therefore, the
increasing of T$_{LT}$ caused by substitution of $^{16}$O by
$^{18}$O for each composition indicates the enhancement of stripe
phase. These results show us a direct evidence for the competing
between static stripe phase and superconductivity. Recently,
Reznik $et$ $al$. found that a strong anomaly in Cu-O
bond-stretching phonon in La$_{2-x}$Sr$_x$CuO$_4$ appears in
superconducting doping level, while disppears in
non-superconducting doping level. It suggests the importance of
electron-phonon coupling to the mechanism of
superconductivity\cite{Reznik}. The anomaly is the strongest in
the samples with static stripe phase:
La$_{1.48}$Nd$_{0.4}$Sr$_{0.12}$CuO$_4$ and
La$_{2-x}$Ba$_x$CuO$_4$. Our isotope substitution confirms the
strong electron-phonon coupling in
La$_{1.6-x}$Nd$_{0.4}$Sr$_{x}$CuO$_4$ system, and supports the
importance of electron-phonon coupling to the mechanism of
superconductivity for high T$_{C}$ superconductors.

In conclusion, oxygen isotope effect is systematically studied in
La$_{1.6-x}$Nd$_{0.4}$Sr$_{x}$CuO$_4$ with static stripe phase.
T$_C$ is suppressed and T$_{LT}$ is enhanced with the substitution
of $^{16}$O by $^{18}$O. These results provide an evidence that
the distortion of CuO$_2$ plane suppresses the superconductivity,
and there exists a competing between static strip phase and
superconductivity. $\alpha$$_C$ shows 1/8 anomaly, similar to that
the observation in Nd free sample. Larger oxygen isotope effect on
T$_C$ is observed compared to the Nd free samples. It indicates
that stronger distortion of CuO$_2$ plane leads to a stronger
electron-phonon coupling. In addition, our results confirm the
strong electron-phonon coupling in the
La$_{1.6-x}$Nd$_{0.4}$Sr$_{x}$CuO$_4$. It is well known that
distortion of CuO$_2$ plane is common feature shared by high-$T_c$
cuprates. Therefore, electron-phonon coupling should play
important role in the mechanism of high-$T_c$ superconductivity

{\bf Acknowledgments:} This work is supported by the grant from
the Nature Science Foundation of China and by the Ministry of
Science and Technology of China (973 project No: 2006CB601001 and
2006CB0L1205).


\begin{thebibliography}{00}

\baselineskip 9pt

\bibitem{Zaanen}
J. Zaanen and O. Gunnarson, Phys. Rev. B {\bf 40} 7391(1989).
\bibitem{Emery1}
V. J. Emery, S. A. Kiveson, O. Zacher, Phys. Rev. B {\bf 56}
6120(1997).
\bibitem{Emery2}
V. J. Emery, S. A. Kivelson adn J. M. Tranquada, Proc. Natl. Acad.
Sci. USA {\bf 96} 8814(1999).
\bibitem{Hayden}
S. M. Hayden $et$ $al$., Phys. Rev. Lett. {\bf 76} 1344(1996).
\bibitem{Yamada1}
K. Yamada $et$ $al$., Phys. Rev. Lett. {\bf 75} 1626(1995).
\bibitem{Yamada2}
K. Yamada $et$ $al$., Phys. Rev. B {\bf 57} 6165(1998).
\bibitem{Tranquada1}
J. M. Tranquada $et$ $al$., Nature {\bf 375}, 561(1995).
\bibitem{Tranquada2}
J. M. Tranquada $et$ $al$., Phys. Rev. B {\bf 54}, 7489(1996)
\bibitem{Qu}
J. F. Qu $et$ $al$., Phys. Rev. B {\bf 71} 094503(2005).
\bibitem{Lee1}
Y. S. Lee $et$ $al$., Phys. Rev. B {\bf 60} 3643(1999).
\bibitem{Moodenbaugh}
A. R. Moodenbaugh $et$ $al$., Phys. Rev. B {\bf 38} 4596(1988).
\bibitem{Homes}
C. C. Homes $et$ $al$., Phys. Rev. Lett. {\bf 96} 257002(2006).
\bibitem{Mook1}
H. A. Mook $et$ $al$., Nature {\bf 395} 580(1998).
\bibitem{Mook2}
H. A. Mook and F. Do$\breve{g}$an, Nature {\bf 401} 145(1999).
\bibitem{Sharma}
R.P. Sharma $et$ $al$., Nature {\bf 404}, 736(2000)
\bibitem{Orenstein}
J. Orenstein and J. Millis, Science {\bf 288} 468(2000) and the
references therein.
\bibitem{Ichikawa}
N. Ichikawa $et$ $al$., Phys. Rev. Lett. {\bf 85} 1738(2000) and
the reference therein.
\bibitem{Axe}
J. D. Axe $et$ $al$., Phys. Rev. Lett. {\bf 62} 2751(1989).
\bibitem{Crawford1}
M. K. Crawford $et$ $al$., Phys. Rev. B {\bf 44} R7749(1991).
\bibitem{Buchner}
B. B\"{u}chner, M. Breuer, A, Freimuth and A. P. Kampf, Phys. Rev.
Lett. {\bf 73} 1841(1994).
\bibitem{Dabrowski}
B. Dabrowski $et$ $al$., Phys. Rev. Lett. {\bf 76} 1348(1996).
\bibitem{Crawford2}
M. K.Crawford $et$ $al$., Phys. Rev. B, {\bf 71} 104513(2005).
\bibitem{Arumugam}
S. Arumugam $et$ $al$., Phys. Rev. Lett. {\bf 88} 247001(2002).
\bibitem{Lanzara}
A. Lanzara et al., Nature {\bf 412}, 510(2001).
\bibitem{Devereaux}
T. P. Devereaux et al., Phys. Rev. Lett. {\bf 93}, 117004(2004).
\bibitem{Crawford3}
M. K. Crawford $et$ $al$., Science {\bf 250} 1390(1990).
\bibitem{Crawford4}
M. K. Crawford $et$ $al$., Phys. Rev. B {\bf 41} 282(1990).
\bibitem{Zhao1}
G. M. Zhao $et$ $al$., Nature{\bf 385} 236(1997).
\bibitem{Zhao2}
G. M. Zhao $et$ $al$., J. Phys.: Condens. Matter {\bf 10}
9055(1998).
\bibitem{Batlogg}
B.Batlogg $et$ $al$., Phys. Rev. Lett. {\bf 59} 912 (1987).
\bibitem{Lee}
Jinho Lee $et$ $al$., Nature {\bf 442} 546 (2006).
\bibitem{Schneider}
T. Schneider and H. Keller, Phys. Rev. Lett. {\bf 86} 4899(2001)
adn the references therein.
\bibitem{Zhao3}
G. M. Zhao $et$ $al$., Nature(London) {\bf 381} 676(1995).
\bibitem{Eisaki}
H. Eisaki et al., Phys. Rev. B {\bf69}, 064512(2004).
\bibitem{Franck}
J. P. Franck, S. Harker and J. H. Brewer, Phys. Rev. Lett. {\bf
71}, 283(1993).
\bibitem{Hofer}
J. Hofer $et$ $al$., Phys. Rev. Lett. {\bf 84}, 4192(2000).
\bibitem{Wang}
G. Y. Wang $et$ $al$., Phys. Rev. B {\bf 74} 165113(2006).
\bibitem{Reznik}
D. Reznik $et$ $al$., Nature(London) {\bf 440}, 1170(2006).
\end{thebibliography}
\end{document}